\documentclass[twocolumn,showpacs,aps,floatfix]{revtex4}

\usepackage{graphicx}
\usepackage{dcolumn}
\usepackage{bm}
\usepackage{amsmath} 
\usepackage{comment} 

\newcommand{\eV}{\,\mathrm{eV}}
\newcommand{\GeV}{\,\mathrm{GeV}}

\newcommand{\Mpc}{\,\mathrm{Mpc}}
\newcommand{\g}{\,\mathrm{g}}
\newcommand{\cm}{\,\mathrm{cm}}
\newcommand{\erg}{\,\mathrm{erg}}
\newcommand{\km}{\,\mathrm{km}}
\newcommand{\kpc}{\,\mathrm{kpc}}
\newcommand{\muG}{\,\mu\mathrm{G}}
\newcommand{\s}{\,\mathrm{s}}
\newcommand{\yr}{\,\mathrm{yr}}

\newcommand{\ud}{\mathrm{d}}
\newcommand{\up}{\mathrm{p}}
\newcommand{\etal}{{\em et al.}}

\newcommand{\tu}[1]{\mathrm{#1}}

\begin{document}

\title{High Energy Neutrinos from Cosmic Ray Interactions in Clusters 
of Galaxies}

\author{Daniel De Marco}
\email{ddm@bartol.udel.edu}
\affiliation{Bartol Research Institute, University of Delaware, Newark, 
DE 19716, USA}
\author{Pasquale Blasi}
\email{blasi@arcetri.astro.it}
\affiliation{INAF/Osservatorio Astrofisico di Arcetri -
Largo E. Fermi, 5 50125 Firenze (Italy)}
\author{Patricia Hansen}
\email{hansen@bartol.udel.edu}
\affiliation{Bartol Research Institute, University of Delaware, Newark, 
DE 19716, USA}
\author{Todor Stanev}
\email{stanev@muon.bartol.udel.edu}
\affiliation{Bartol Research Institute, University of Delaware, Newark, 
DE 19716, USA}

\date{\today}

\begin{abstract}
The spatial clustering of galaxies in galaxy clusters implies that the
background of infrared (IR) light in the intracluster medium (ICM) 
may exceed the universal background. Cosmic rays injected within the ICM
propagate diffusively and at low enough energies are trapped there for 
cosmological times. The photopion production interactions of cosmic rays 
with the IR photons are responsible for the generation of neutrinos 
whose detection may shed some light on the origin and propagation of high 
energy cosmic rays in the universe. Here we discuss our calculations of 
the flux of neutrinos from single clusters as well as the contribution 
of photopion production in clusters of galaxies to the diffuse neutrino 
background. 
\end{abstract}

\maketitle

\section{Introduction}

Clusters of galaxies are peculiar storage rooms of cosmic rays 
\cite{bbp97,volk96}: most non-thermal protons and nuclei injected in 
the ICM remain confined there for cosmological times, and can either 
interact with the gas and photon backgrounds or be re-energized 
by episodic phenomena, such as mergers with other clusters 
\cite{sarazin,blasi01}. The spatial clustering of 10-1000 galaxies within
the cluster volume also enhances the amount of infrared and optical 
light in that region, thereby increasing the probability of interactions
of charged particles with these photons. The enhanced cosmic ray 
and photon densities make clusters of galaxies the ideal laboratory
for the generation of high energy radiations and neutrinos. The detection
of a neutrino flux from clusters of galaxies would be a precious 
tool to weigh the unknown non-thermal content of these large scale
structures. As discussed in Sec. \ref{sec:cr}, the amount of cosmic 
rays trapped in the intracluster medium is still very uncertain, and
only weakly constrained by present observations of hard X-rays and 
gamma rays, most likely generated by high energy electrons 
\cite{brunettirev,blasirev}. The detection of neutrinos on the other 
hand would allow us to have an independent handle to infer the energetic 
non-thermal content which is in the form of hadronic cosmic rays. 
Gamma rays in the 1-1000 GeV range are also expected to carry precious 
information about the amount of cosmic rays trapped in the ICM
\cite{cb98,ensslin}.

In this paper we investigate the most promising mechanism 
for the generation of high energy neutrinos, namely the production 
and decay of charged pions in proton-photon inelastic interactions.
The relevant photon background in the ICM is provided by the IR and
optical light emitted by galaxies within the cluster. The maximum
flux of neutrinos from inelastic $pp$ interactions allowed by the 
observations of the diffuse gamma ray background was calculated in
\cite{bbp97,cb98}. The calculation of the neutrino flux is carried out 
both analytically and through a Monte Carlo simulation of the 
diffusive motion of high energy particles in the turbulent magnetic 
field of a cluster of galaxies.
Such propagation was also investigated in \cite{rodo} with the purpose
of determining the flux of gamma and hard X-rays generated as secondaries
of the high energy protons.

The paper is organized as follows: in Sec. \ref{sec:cr} we review
the current knowledge of the cosmic ray propagation and energetics 
in the ICM. In Sec. \ref{sec:IR} we illustrate our calculations of
the IR light background and its spatial distribution in the 
volume of the cluster. In Sec. \ref{sec:simple} we outline
a benchmark analytical calculation of the neutrino flux, in order to 
emphasize the dependence of the result upon poorly known quantities,
such as the diffusion coefficient and the CR energy content of the 
cluster. A Monte Carlo calculation of these neutrino fluxes is 
described in Sec. \ref{sec:MC}. In Sec. \ref{sec:diffuse} we calculate
the maximum expected contribution of these processes of neutrino production
in clusters of galaxies to the diffuse extragalactic neutrino flux. 
We conclude in Sec. \ref{sec:last}.

\section{Cosmic Rays in the intracluster medium}\label{sec:cr}

The presence of a non-thermal component in clusters of galaxies
is best shown by the observation of regions of extended radio emission,
called {\it radio halos}, with size 1--2~Mpc, often wider than the
X-ray emitting region (see \cite{feretti} for a recent review). 
The spectrum of the radio emission is 
typically a power law with a high frequency cutoff in the 1--10~GHz
range and it results from synchrotron energy losses of high 
energy electrons in the magnetic field of the ICM. Measurements 
of the rotation measure provide magnetic fields which are of the
order of several $\muG$ \cite{kim,clarke}, although several of these 
inferred values depend on assumptions made on the structure of the magnetic 
field and on the density of background electrons. In a few clusters, 
the radio emission is accompanied by X-rays in excess of those of 
thermal origin (due to bremmstrahlung emission). These non-thermal 
X-rays are most likely the result of inverse Compton scattering (ICS) 
of the same electron population responsible for the radio emission. 
The main photon target is represented by the photons of the 
cosmic microwave background (CMB), while the infrared light, even 
in the case of the enhanced fluxes considered here, does not play a 
significant role. In the few cases in which both a radio 
halo emission and a hard X-ray emission are detected, an estimate
of the magnetic field can be obtained \cite{fusco}. Typical values are in the
range 0.1--1$\muG$. These fields should be considered as averages over
the volume of the cluster. Higher fields are likely to be present in
the central regions of clusters.

While non-thermal electrons are easily {\it visible} in clusters,
a much harder task is to identify hadronic cosmic rays. 
The smoking gun showing their presence, namely a gamma ray emission 
associated with the decay of neutral pions, is still lacking. This 
absence of detection is however still compatible with most models of 
injection of cosmic ray protons in the ICM: the expected gamma
ray emission becomes comparable with the EGRET sensitivity only
if the protons in clusters have a total energy comparable
with the thermal energy in the virialized gas. Lower fluxes
should however be detectable with the upcoming GLAST satellite
(\cite{gabici2,gabici3}, see also \cite{blasirev} for a recent review). 
One of the main reasons for the interest in the cosmic ray
population in clusters is related to the fact that the presence
of magnetic fields is able to confine cosmic rays for times
in excess of the age of the cluster: these large scale structures
work then as storage rooms for cosmic rays \cite{bbp97,volk96}. 
This implies that the energy density of cosmic rays in clusters 
increases with cosmic time.
The potential sources of cosmic rays in clusters have been 
considered in \cite{bbp97}  and the corresponding
energy densities were estimated. The efficiency of the confinement
was also investigated: the maximum energy for which the confinement
is effective, namely the confinement time exceeds the age of 
the cluster (roughly the age of the universe), is strongly 
dependent upon the energy dependence of the diffusion coefficient.
For Bohm diffusion and $\muG$ fields, the confinement remains 
effective up to extremely high energies, while a Kolmogorov spectrum 
of magnetic fluctuations leads to confinement only up to energies of the 
order of several TeV. 

Normal galaxies in clusters are expected to contribute a total 
energy in the form of confined cosmic rays $\epsilon \approx 
L_\up N_\tu{gal} t_0$, where $N_\tu{gal}\sim 100$ is the typical number 
of galaxies in a cluster, $L_\up\sim 3\times 10^{40}\rm erg~s^{-1}$ 
is the cosmic ray luminosity of a galaxy like ours. We then obtain 
$\epsilon\approx 10^{60}$ erg, more than three orders of magnitude 
less than the total energy in the thermal gas. A single active galaxy in 
a cluster should easily contribute a cosmic ray luminosity of 
$\sim 10^{44}\rm erg~s^{-1}$.
At any time it is reasonable to assume that at least one of the
galaxies in the cluster is in an active phase, so that the total energy 
contributed by active galaxies is expected to be $\sim 3\times 
10^{61}\rm erg$. A similar estimate is obtained if cosmic rays
are accelerated through first order Fermi acceleration at the 
accretion shock in the outskirts of the cluster, if the acceleration
efficiency is $\sim 10\%$. These shocks form in the outskirts of 
clusters and are the result of the propagation of the information
on the virialization of the core towards the outer regions \cite{bert}. 
Accretion shocks propagate in the cold intergalactic gas and
therefore their Mach number can be very high. In the context
of the linear theory of particle acceleration at non-relativistic shocks,
the expected spectrum of accelerated particles in the limit of very 
large Mach numbers is a power law with slope $\alpha\approx 2$. 
The cosmic ray induced modification of the shock, particularly 
important for high Mach numbers, may generate higher efficiencies 
and non-power-law spectra \cite{blasirev,stefano,jones}. 

Shock waves are also expected to appear due to the supersonic infall
of two or more clusters in a merger event. The typical strength of
these shocks is much lower than for accretion shocks, and the 
expected spectra of accelerated particles is steep \cite{gabici1,dermer}.
The energy liberated in the gravitational form during these events 
is of order $\sim GM_\tu{DM}^2/R_\tu{vir}\sim 10^{64}\rm erg$, which is
in fact of the same order of magnitude of the thermal energy in the
cluster, as should be expected since these mergers are believed to be
responsible for the heating of the ICM to its virial value. Some
fraction of this energy may be in the form of accelerated protons, but
the efficiency for particle acceleration is expected to be rather small
because of the low Mach numbers, which are also responsible for steep
spectra \cite{gabici1}.

From the observational point of view, the most severe constraint
to the energy density of cosmic rays, though model dependent, is
represented by the radio emissivity at a few GHz frequency. 
Radio emission at this frequency is generated by electrons
with energy $E_e = 37 \,\rm GeV\: B_\mu^{1/2}$, where $B_\mu$ is the
magnetic field in $\muG$. Observations of the
Coma cluster show a cutoff in the volume integrated radio spectrum, 
at $\nu\sim 5 \,\rm GHz$ \cite{observa}. 
If cosmic rays are present, a contribution to the 
radiating electrons comes from secondary electrons, generated 
from $pp$ inelastic collisions. The spectrum of secondary electrons 
however has no cutoff at such low energies, therefore the
radio emissivity at high frequencies imposes a rather severe
constraint on the amount of secondary
electrons, and therefore of primary cosmic rays in the region
$10-1000\GeV$ \cite{Reimer2004}. This bound implies that 
cosmic rays amount to less than $(1-10)\%$ of the thermal energy
 in the cluster,
depending upon assumptions on the spectrum at injection and the
morphology of the magnetic field in the cluster. These limits are
slightly tighter if reacceleration of the secondary electron-positron
pairs is taken into account \cite{brunetti}.

\section{The infrared light in the intracluster medium}\label{sec:IR}

In this section we illustrate our determination of the InfraRed
(IR) background in the intracluster medium. We start with adopting the 
Spectral Energy Distribution (SED) of individual galaxies following
\cite{LDP03}. We then carry out the convolution of these SEDs
with the spatial distribution of galaxies in a sample cluster of
galaxies. 

\begin{figure}[thb] 
\includegraphics[width=85truemm]{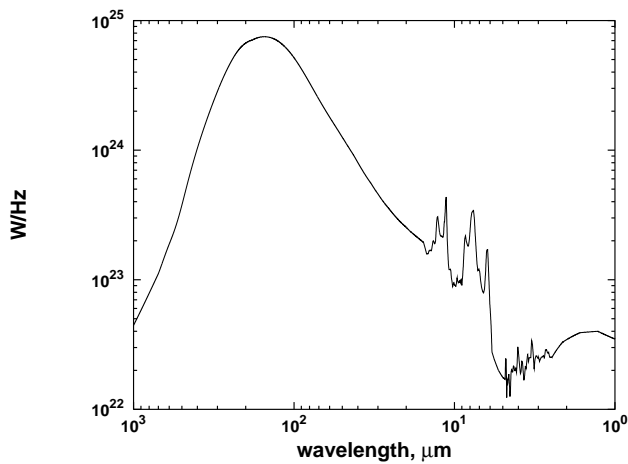}
\caption{Model spectrum of a normal galaxy of luminosity
\protect$10^{11} L_\odot$~\cite{LDP03,LDPweb}.}
\label{fig:cold1}
\end{figure}

The SED of a $10^{11} L_\odot$ normal galaxy as obtained by 
\cite{LDP03,LDPweb} is illustrated in Fig.~\ref{fig:cold1}. 
The authors of Ref.~\cite{LDP03,LDPweb} state that the SED they find 
is valid only at wavelengths longer than 4$\mu$m~\cite{LDP03}, but 
after comparing it with other models and data we decided to use it 
at longer wavelengths as well, since the differences were found to 
be smaller than any other uncertainty of astrophysical origin. The 
same comment holds for the differences due to different types of galaxies 
in their model. 

We used 1$\mu$m as upper energy for the background radiation since at
higher energies its spectrum is much steeper and even with a steep
proton spectrum there is no appreciable contribution to the neutrino
production.

\begin{figure}[thb] 
\includegraphics[width=85truemm]{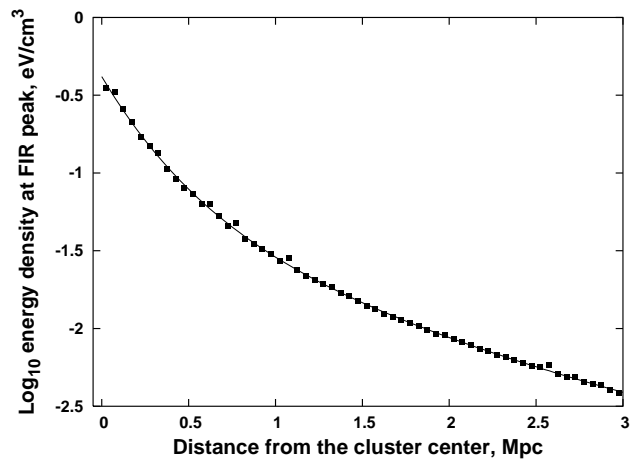}
\caption{Energy density at the FIR peak for a particular realization
(points) and its fit (line) that can be used for other realizations.}
\label{fig:irbg1}
\end{figure}

\begin{figure}[thb] 
\includegraphics[width=85truemm]{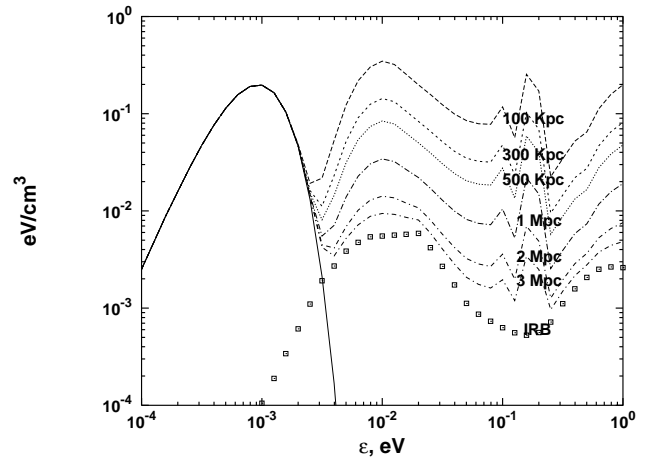}
\caption{Energy density of the IR (cluster + extragalactic)
for six zones inside the cluster. Points show the extragalactic 
background from Ref.~\cite{F01}.}
\label{fig:mbrpl2}
\end{figure}

In order to determine the IR background within the cluster of galaxies,
we calculate the convolution of the given SED from our {\it prototypical}
galaxy with the spatial distribution of galaxies. Following \cite{rodo}
we assume that the galaxy distribution follows closely the distribution of
gas, for which we use:
\begin{equation}
  f(r)\propto\Big[
  \Big(1+\frac{r}{r_1}\Big)^{0.51}
  \Big(1+\frac{r}{r_2}\Big)^{0.72}
  \Big(1+\frac{r}{r_3}\Big)^{0.58}
  \Big]^{-1},
\end{equation}
with $r_1=10\kpc$, $r_2=250\kpc$ and $r_3=1\Mpc$.
Fig.~\ref{fig:irbg1} shows the energy density at the peak of the SED 
as a function of the distance from the center of the galaxy cluster,
in the assumption that $N_g\sim 1000$ galaxies are in the cluster. 
The function shown in Fig. \ref{fig:irbg1} can be used to construct 
a useful model of the IR density as a function of the distance
to the center of the cluster by integration over different distance 
ranges. This is shown in Fig. \ref{fig:mbrpl2} where the sum of the 
IR emission from cluster sources is added to the extragalactic IR 
background and plotted in six circular annuli inside the cluster.

\section{Neutrino production: an analytical
calculation}\label{sec:simple}

In this section we present an analytical estimate of the neutrino 
flux generated as a result of the photopion interactions of high 
energy protons with the IR and CMB photons in the ICM. For simplicity we
assume that there is only one dominant source of energetic protons in the 
center of the cluster and that the IR photon background is constant
within the inner $500\kpc$ of the cluster. We also assume that 
the diffusion coefficient within this region is spatially constant.  

Under these assumptions the neutrino production in the cluster can be
calculated as: 
\begin{align}\label{eq:nuinj}
  \nonumber j_\nu(E_\nu) = \int_{E_\up^\tu{min}}^{E_\up^\tu{max}}
 \ud E_\up & \int_{0}^{r_\tu{diff}(E)} \ud r 
        \frac{\ud n_\nu}{\ud E_\nu}(E_\up,E_\nu)\times\\
        &\times n_\up(E_\up,r)\,
        \frac{c}{\ell(E_\up)}\,
        4\pi r^2\,
        \,,
\end{align}
where $\ud n_\nu/\ud E_\nu$ is the average spectrum of neutrinos
produced by a single proton interaction on the photon background,
$n_\up(E_\up,r)$ is the number of protons at distance $r$ from the
center and $c/\ell(E_\up)$ is the interaction probability per unit time.

The number density of protons at distance $r$ from the central source,
$n_\up(E,r)$, can be easily determined by solving the diffusion equation
\cite{bbp97}, obtaining:
\begin{equation}
n_\up(E,r)=\frac{Q(E)}{4\pi r D(E)}.
\label{eq:equi}
\end{equation}
This solution is valid for particles that in the time $t_0$ manage
to diffuse out to a distance $r$ from the center without suffering
appreciable energy losses. This means particles whose energies
satisfy $r^2/(4 D(E)) < t_0$. At distances larger than $\sqrt{4 D(E)
t_0}$ the density of particles with energy $E$ drops exponentially. 

Here we assume that the diffusion coefficient has the form found 
in \cite{parizot}:
\begin{equation}
D(E)=D_* \left[ \left(\frac{E}{E_*}\right)^{1/3} +
\left(\frac{E}{E_*}\right) +
\left(\frac{E}{E_*}\right)^2 \right],
\label{eq:diffusion}
\end{equation}
where $D_*=\frac{1}{4}r_\tu{L}(E_*) c$, $r_\tu{L}(E_*)$ is the Larmor radius of
particles with energy $E_*$ and the reference energy $E_*$ is found by
requiring that $r_\tu{L}(E_*)=L_\tu{c}/5$, $L_\tu{c}$ being the coherence scale of
the magnetic field. The three terms in Eq. \ref{eq:diffusion}
correspond to different regimes of propagation (Kolmogorov, Bohm and
pitch angle scattering) depending upon the comparison of the Larmor
radius of the particle and the coherence scale of the magnetic field.

The upper limit in the integral over the radial coordinate in Eq.~\ref{eq:nuinj}
is defined as $r_\tu{diff}(E)=\min(\sqrt{4D(E)t_0},r_\tu{max})$, with
$r_\tu{max}=500\kpc$ and $t_0$ the age of the cluster, comparable with
the age of the universe, $t_0 \sim 10^{10}\yr$. It is easy to show,
following \cite{bbp97}, that, for cosmic ray energies for which
confinement works, the spectrum of secondaries (gamma rays and
neutrinos) is not affected by diffusion and is therefore not steepened
by the energy dependence of $D(E)$. For unconfined cosmic rays,
diffusion plays a role and the spectrum of secondaries suffers a
steepening. 

\begin{figure}
  \includegraphics[width=85truemm]{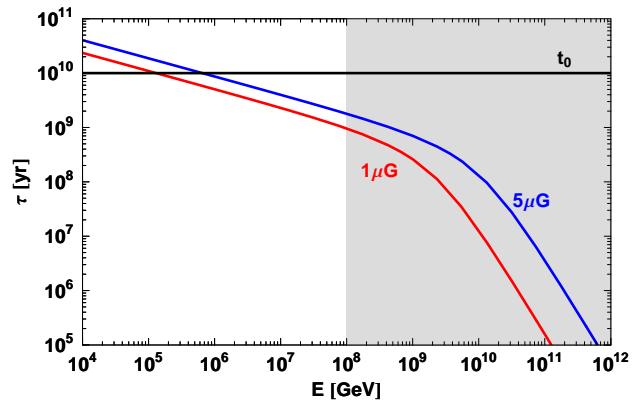}
  \caption{Diffusion time as a function of the proton energy. In the
  energy region considered here, the gray region in the plot, the
  diffusion time is always shorter than the age of the cluster.}
\label{fig:rmax}
\end{figure}

For a given diffusion coefficient, the minimum energy of the particles
for which the escape time does not exceed $t_0$ can be estimated by
requiring that $r_\tu{max}^2/4D(E)<t_0$. In Fig.~\ref{fig:rmax} we plot
the diffusion time, $\tau(E)=\frac{r^2_\tu{max}}{4 D(E)}$, out of a sphere
with radius $r_\tu{max}$ as a function of energy, using as a diffusion
coefficient the one reported in Eq.~\ref{eq:diffusion}. The two curves
correspond to magnetic fields in the cluster $B_0=1\muG$ (lower curve)
and $B_0=5\muG$ (higher curve). As it is clear from the plot, due to the
large energy threshold for photopion reaction, all the particles of
interest for our problem are not confined on time scales of the order of
the age of the cluster. It is however worth keeping in mind that this
estimate is rather simplified and that more realistic distributions of
magnetic field may reflect into longer confinement times as we actually
find in Sec.~\ref{sec:MC}. 

We used the SOPHIA event generator \cite{sophia} to calculate the
average spectrum of neutrinos produced in a single proton interaction
on the cluster photon background and we parametrized the results as:
\begin{equation}
\frac{\ud n_\nu}{\ud E_\nu}(E_\up,E_\nu)=
\frac{45}{E_\up} \exp\Big[
-\Big(\frac{E_\nu}{E_\up/20}\Big)^{1.2}\Big]\,,
\end{equation}
for $E_\nu<E_\up$ and 0 otherwise. 
In order to calculate the interaction probability per unit time we use the
interaction lengths as calculated in the next section and we
approximate the interaction length in the region with $r<500\kpc$ as:
$\ell(E_\up)\simeq3\cdot10^{14}\big(\frac{E}{E_0}\big)^{-1.2}\Mpc$, with
$E_0=1\GeV$ (see Fig.~\ref{fig:alampg1}).

The integration limits for the proton energy in Eq. \ref{eq:nuinj}
are: $E_\up^\tu{max}=10^{11}\GeV$ and $E_\up^\tu{min}=\max(E_\nu, 
E_\tu{th}=10^{8}\GeV)$ (for $E_\up<10^8\GeV$ the neutrino production is
negligible).

For the purpose of numerical calculations, we assume, following
\cite{bbp97,bc}, that the cosmic ray source has a luminosity, above 
energy $E_0=1\GeV$, $L_\up=3\times 10^{44}\erg\s^{-1}$ (see also Sec.
\ref{sec:cr}).

\begin{figure}
  \centering
  \includegraphics[width=85truemm]{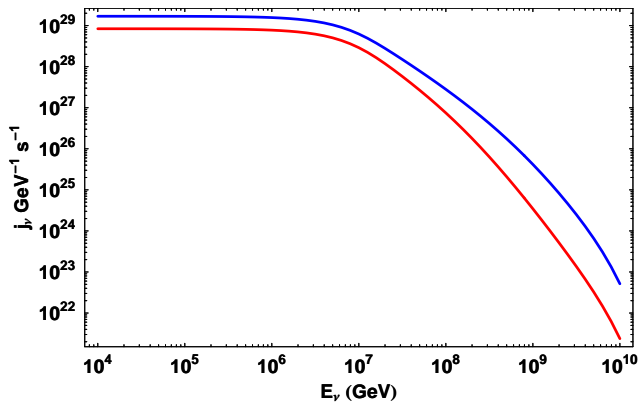}
  \caption{Analytical estimate of the neutrino emissivity from 
  proton propagation in a cluster, with slope of the injection spectrum
  of protons $\alpha=2.1$. Upper curve for $B_0=5\muG$, lower curve for
  $B_0=1\muG$.}\label{fig:nusp}
\end{figure}

The neutrino spectrum from Eq. \ref{eq:nuinj} for $\alpha=2.1$ is plotted in
Fig.~\ref{fig:nusp} for magnetic field $1\muG$ (lower curve) and 
$5\muG$ (upper curve). 
Assuming that the cluster is located at a typical distance of
$100\Mpc$, our estimate of the neutrino flux at the Earth is of 
$0.29\km^{-2}\yr^{-1}$ for magnetic field $1\muG$ and 
$0.7\km^{-2}\yr^{-1}$ for magnetic field $5\muG$.

This result varies somewhat with different choices of the diffusion
coefficient, but not in a substantial way. Steeper injection spectra
of cosmic rays imply smaller neutrino fluxes. 

\section{Neutrino production: a Monte Carlo approach}\label{sec:MC}

The result obtained in the previous section may somewhat be affected  
by the adoption of a more detailed distribution of the IR photons in 
the ICM, as well as by details of the diffusion in a space dependent 
magnetic field. 
In particular, both the IR light and the magnetic field are expected to
be higher in the central region of the cluster. In order to
investigate the effect of these factors on the predicted neutrino
flux, we performed a Monte Carlo calculation of the propagation and
the related neutrino production. 

\begin{figure*}
  \centering
  \includegraphics[width=140truemm]{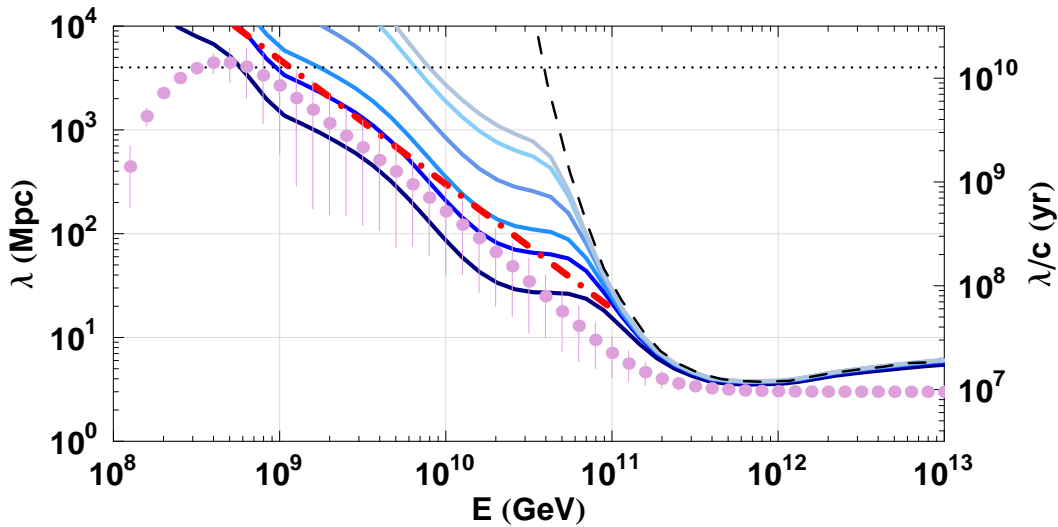}
  \caption{Interaction lengths of protons propagating inside the six
  concentric zones in the cluster. The solid lines correspond, from
  bottom to top, to the zones with radii: $100\kpc$, $300\kpc$,
  $500\kpc$, $1\Mpc$, $2\Mpc$, $3\Mpc$. The dashed line is the
  interaction length on the CMB. The dotted line represents the size of
  the universe. The dot-dashed line represents the approximate
  interaction length we used in the analytical calculation. Data points
  show the containment distance in the cluster for central magnetic
  field of 5\protect$\mu$G.}\label{fig:alampg1}
\end{figure*}

The photon background is taken as the superposition of the IR
background, as calculated in Sec.~\ref{sec:IR}, and the usual CMB. 
We divide the cluster in six concentric zones with radii: $100\kpc$, 
$300\kpc$, $500\kpc$, $1\Mpc$, $2\Mpc$ and $3\Mpc$. In each zone we 
assume the infrared background to be constant and equal to the 
average one in that zone. The interaction lengths calculated in 
each zone are plotted in Fig.~\ref{fig:alampg1}.

The simulation of the propagation of charged particles in the ICM
requires a specific choice of the strength, spatial profile and
disordered component of the magnetic field. Inspired by the results of
\cite{mf1,mf2} on the cluster Abell 119, we assume that the total
magnetic field scales with distance from the center of the cluster as 
given by the flux freezing condition with an electron thermal
component modelled as a $\beta$-model:
\begin{equation}
  B(r)\propto\Big(1+\frac{r^2}{r_c^2}\Big)^{-0.7}
\label{eq:br}
\end{equation}
with $r_c=378\kpc$. This result agrees quite well with both the theoretical
expectation of $B(r)\propto n_\tu{e}(r)^{2/3}$ for a magnetic field frozen
in matter ($n_\tu{e}(r)$ is the electron density) and with the
simulation results of Ref.~\cite{rodo} (see their Fig.~1).
As a result, the average magnetic field strength in the six zones of the
cluster changes from about 4.5 $\mu$G in the inner zone to about 0.18 $\mu$G
in the outer zone in the case of 5 $\mu$G strength in the center. The
average field values are scaled down by a factor 5 for the 1 $\mu$G case. 
The simulation code does not use these averages, rather the magnetic field
at the particle location.

We implemented a turbulent magnetic field with the above radial
dependence dividing the ICM in cubes of $50\kpc$ side, each filled
with a random magnetic field of average strength $\langle B\rangle =B_0$
satisfying a Kolmogorov spectrum with three logarithmic sub-scales
($12.5, 25, 50\kpc$). Given the proton position we calculate the
turbulent magnetic field in the cube containing the particle and 
then we scale its magnetic field according to Eq.~\eqref{eq:br}. 
For more informations and details about the actual implementation 
see Appendix B of Ref.~\cite{simu}.

We calculate the neutrino yield, $Y(E_\up, E_\nu)$, namely the 
spectrum of neutrinos generated during the propagation of protons 
of energy $E_\up$ in the ICM. This calculation is carried out
by injecting protons in 30
logarithmic energy bins between $10^{17}\eV$ and $10^{20}\eV$.
We inject 10,000 protons for each energy bin.
We follow their propagation in the ICM until their distance from the
center exceeds $3\Mpc$ or their energy falls below $10^{17}\eV$.
Clearly the neutrino yield as defined here depends on the
specific model for injection of the particles. We limit ourselves to
considering only the case of a source that generates cosmic rays in 
the center of the cluster.

The propagation is simulated using the Monte-Carlo code described in 
Ref.~\cite{simu}. The nucleon-photon interactions are simulated with 
the event generator SOPHIA~\cite{sophia}. At each interaction we
record the secondary products and then at the end of the simulation 
we have the spectrum of the neutrinos produced inside the cluster by 
a proton injected with a given energy, $Y(E_\up, E_\nu)$. 
 
We ran our simulation for two values of the magnetic field in the
center of the cluster, $B_0=1\muG$ and $B_0=5\muG$. In
Fig.~\ref{fig:alampg1} we plot the pathlength travelled by particles 
propagating in the cluster as a function of energy
for the case of injection in the center and for $B_0=5\muG$. 
For each energy bin we 
calculate the average pathlength travelled. In this average we include 
particles that exit the cluster and particles that go below
threshold. This considerably reduces the pathlengths at low energy 
(below $10^{18}\eV$). 

\begin{figure}
  \centering
  \includegraphics[width=85truemm]{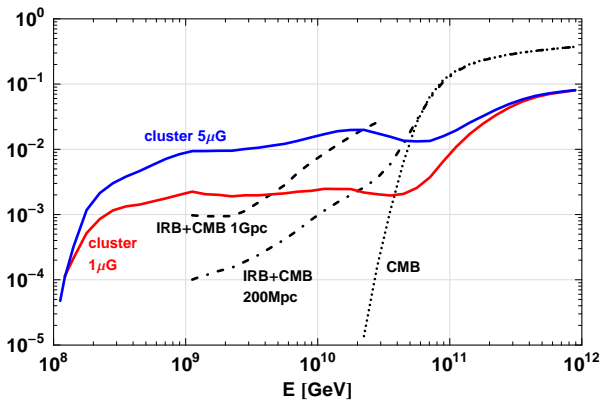}
  \caption{The fraction of the proton energy converted to neutrinos inside 
  the cluster is shown with solid lines for 5\protect$\mu$G and
  1\protect$\mu$G central fields. The dotted curve shows the energy
  in neutrinos for propagation in the CMB on 200~Mpc. The dash-dotted
  and dashed lines show the fraction of energy in neutrinos for
  propagation in the cosmic IR background on scales of 200~Mpc and
  1~Gpc.}\label{fig:nufrac} 
\end{figure}

\begin{figure}
  \centering
  \includegraphics[width=85truemm]{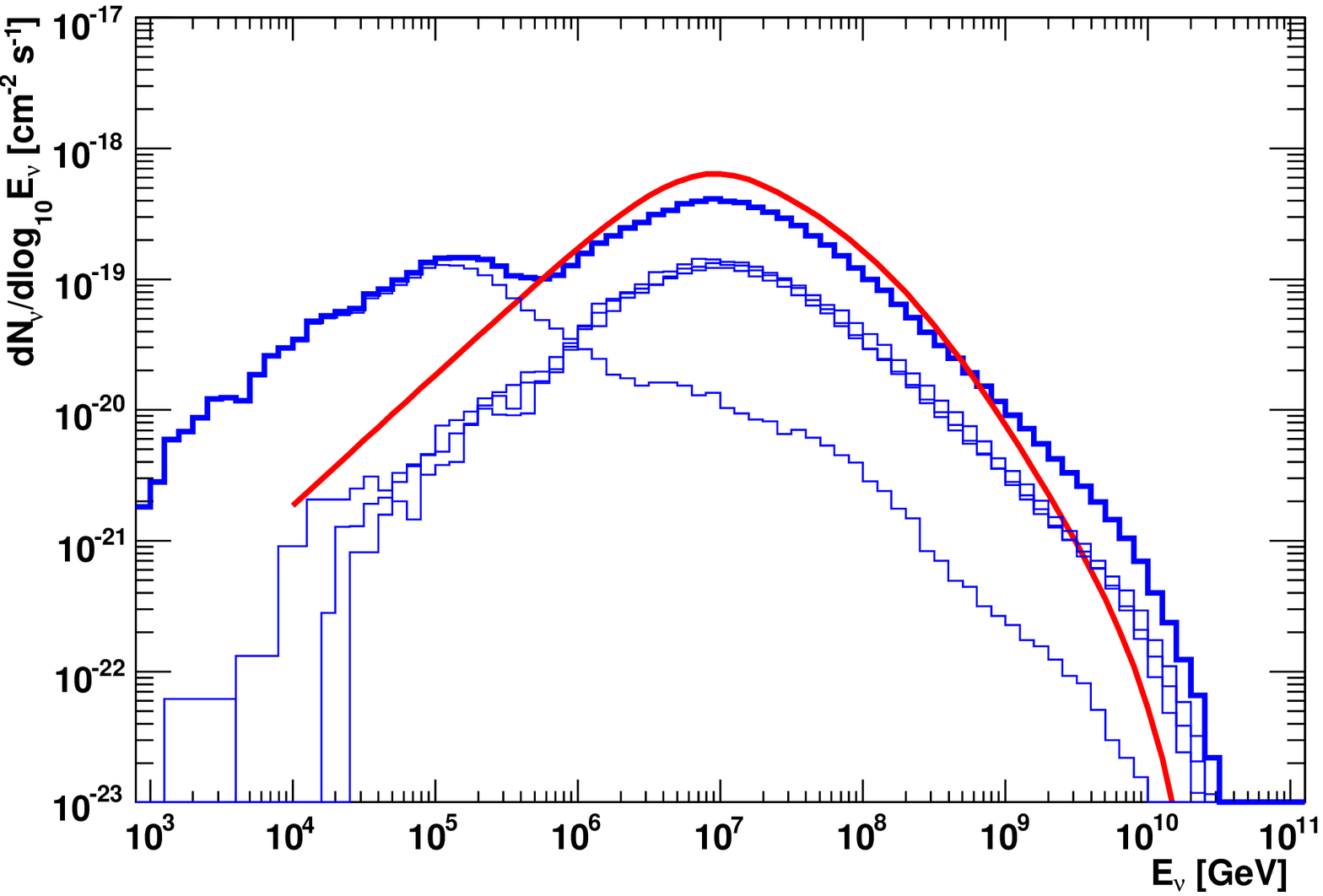}
  \includegraphics[width=85truemm]{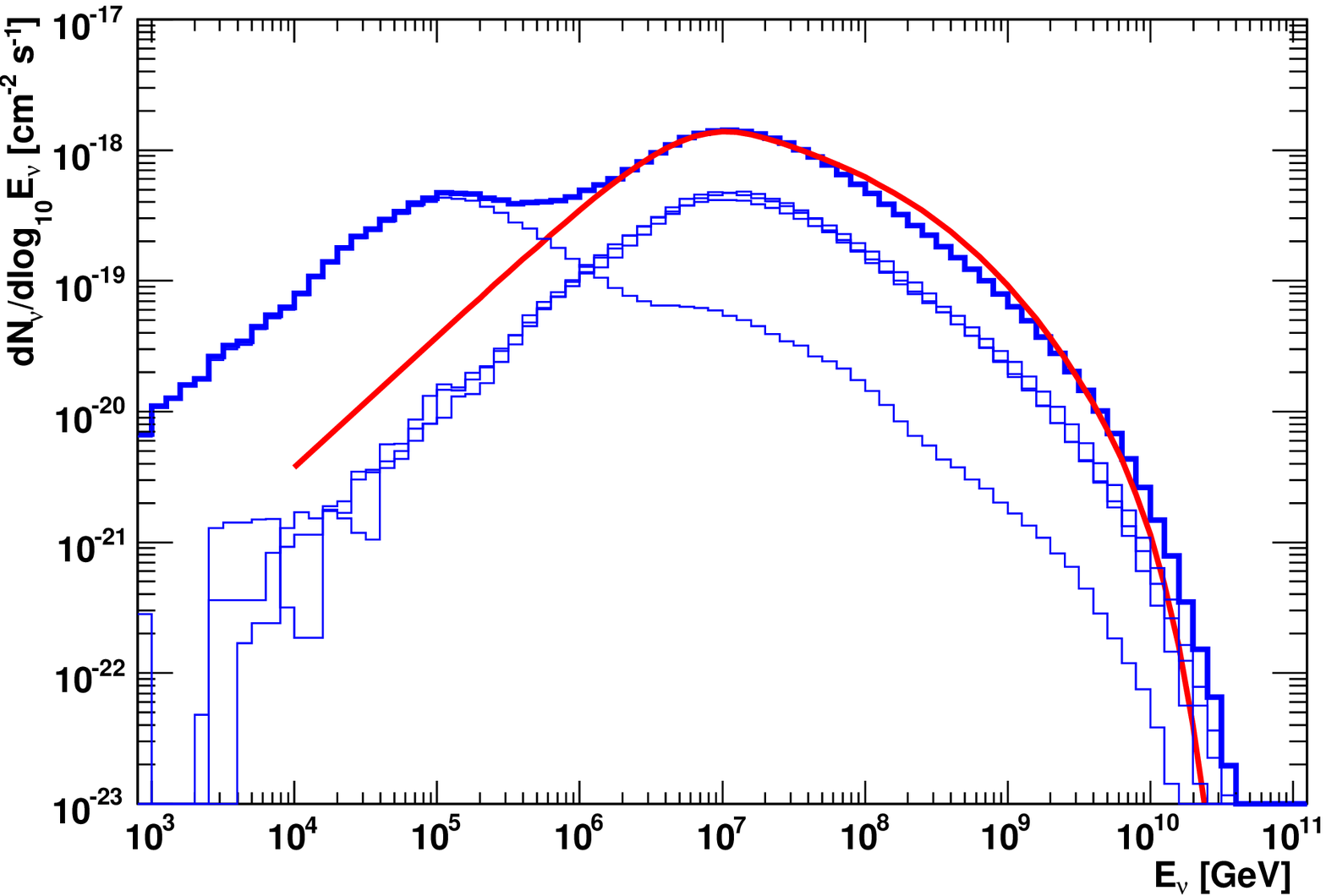}
  \caption{Flux of neutrinos from a cluster located at a typical
  distance of 100 Mpc from the Earth. The upper (lower) panel refers to
  the case $B_0=1\muG$ ($B_0=5\muG$). The different histograms show the
  flux of neutrinos of different flavors, while the thick histogram
  shows the total flux. The continuous line is the result of the
  analytical estimate in Sec.~\ref{sec:simple}.}
\label{fig:res}
\end{figure}

Since at about $\sim 10^{17}\eV$ the interaction time becomes 
larger than the age of the universe (see Fig.~\ref{fig:alampg1})
we decided to use only the protons with energy above $10^{17}\eV$ to
calculate the neutrino flux. Lower energy protons, even with a steep
spectrum, do not change the predicted neutrino flux because at these
energies they lose their energy mainly due to adiabatic energy losses.

Fig.~\ref{fig:nufrac} shows the fraction of the proton energy that is
converted to neutrinos before leaving the cluster. For energies around
2$\times$10$^{17}$ eV this fraction is 0.1\% and almost linearly increases
with $\ln E_p$ to about 1\% at 10$^{18}$ eV and stays roughly 
constant at higher energy. The proton energy loss in the cluster 
is a function of the hadronic cross section and of the containment 
time in the cluster.
Because protons of energy approaching $10^{20}$ eV propagate almost 
in straight lines inside the cluster they are not expected to be 
very efficient in generating neutrinos.  

For the case of protons injected at the center of the cluster, the 
neutrino spectrum is given by
\begin{equation}
  j^i_\nu(E_\nu) = \int Y^i(E_\up,E_\nu)\,Q_\up(E_\up) \ud E_\up,
\end{equation}
where: $Y^i(E_\up,E_\nu)$ is the neutrino yield, for neutrinos of type
$i$, at given proton energy and $Q_\up(E_\up)$ is
the injection rate of protons in the cluster. 

The results of our Monte Carlo calculations are shown in
Fig. \ref{fig:res}. The upper (lower) panel refers to the case
$B_0=1\muG$ ($B_0=5\muG$). Both panels show the well known
pronounced two peaks \cite{ESS01} due to the direct neutrino production
(higher energy peak) and neutrino production following neutron decay
(lower energy peak).
 The $\bar{\nu}_e$ from neutrons decaying outside the cluster are
 not included.
 The fluxes of neutrinos of different flavors are
superimposed and the thick histogram represents the total flux. The
continuous line is the result of the analytical calculations
discussed in Sec.~\ref{sec:simple}. The agreement between the
analytical and Monte Carlo calculations is remarkable, although the 
former did not include the decay of neutrons, and therefore has no low
energy peak.

The integrated flux of neutrinos with energy $E>10^5$ GeV as
calculated from the Monte Carlo for a cluster at a typical distance of
100 Mpc is $0.9 \; \rm km^{-2} yr^{-1}$ for $B_0=5\muG$ and $0.2 \; \rm
km^{-2} yr^{-1}$ for $B_0=1\muG$. One should keep in mind that this
flux is obtained for a source luminosity in the form of protons  
$L_\up=3\cdot10^{44}\erg\s^{-1}$. The neutrino flux scales linearly
with $L_\up$, and may be appreciably higher than those predicted here
if a substantially more luminous source happens to be located in a
nearby cluster. On the other hand, the neutrino flux is dominated by 
neutrinos with energies around $10^{17}$ eV, generated by particles that 
travel within the ICM for a substantial fraction of the age of the
cluster. The luminosity reported above should therefore be interpreted
as the proton luminosity averaged over a period comparable with the
age of the cluster and it is
unlikely to find values of $L_\up$ much higher than that used here. 

\section{The Diffuse Neutrino Flux}\label{sec:diffuse}

In this section we illustrate our calculations of the contribution 
of photopion production in clusters of galaxies to the diffuse 
neutrino flux. For this purpose, we use the results of our Monte 
Carlo calculations (Sec.~\ref{sec:MC}) to obtain a template spectrum 
to be convolved with the mass distribution of clusters of galaxies. 
For the integral mass function of clusters of galaxies we use the 
following expression \cite{bc93}:
\begin{equation}
n(>M) = 4\times 10^{-5} \left( \frac{M}{M_*} \right)^{-1} 
\exp \left(-\frac{M}{M_*}\right) h^3 \Mpc^{-3},
\end{equation}
where $M$ is the total (gravitational) mass of a cluster within $1.5\,h^{-1}\Mpc$ 
and $M_* = (1.8 \pm 0.3)\times 10^{14}\,h^{-1} M_\odot$ is a reference
cluster mass. This distribution covers the range from $M=10^{12} h^{-1}
M_\odot$ to $M=5\times10^{15} h^{-1} M_\odot$.
For our numerical calculations we adopt $h=0.7$ for the
dimensionless Hubble constant. The flux of diffuse neutrino radiation
from clusters of galaxies, as due to photopion production can be written
as
\begin{equation}
\Phi(E_\nu) \approx \frac{c\,t_0}{4\pi} \int \ud M \frac{\ud n}{\ud M} J_\nu(E_\nu,M), 
\label{eq:diffuse}
\end{equation}
where we neglect the effect of redshift since the uncertainties in 
the determination of the neutrino emissivity, $J_\nu$, are very large compared
with the effects of cosmology. Rather than presenting several cases of diffuse
neutrino fluxes, depending on the choice of the several parameters that enter
the calculations, we decided here to show the maximum neutrino flux. In 
order to estimate this flux, we assume that the maximum allowed cosmic ray  
luminosity in the cluster is 
\begin{align}
L_\up  = \xi \frac{G M^2}{R_\tu{v} t_\tu{cl}} &= 3\times 10^{45}
\left(\frac{\xi}{0.1}\right)\times\nonumber \\
&\times \left(\frac{M}{5\times 10^{14} \, M_\odot}\right)^{5/3} 
\erg\s^{-1},
\label{eq:lumi}
\end{align}
where we used the definition of virial radius as
\begin{equation}
R_\tu{v} = \left(\frac{3M}{4\pi \Delta_c \Omega_\tu{M} \rho_\tu{cr}}\right)^{1/3},
\end{equation}
with $\Delta_c\approx 200$, $\Omega_\tu{M}=0.3$ for the matter fraction in the
universe and $\rho_\tu{cr}=1.88\times 10^{-29}\,h^2\g\cm^{-3}$ for the 
critical density. One should keep in mind that luminosities of this order
of magnitude can only be reached in the context of major mergers of clusters 
of galaxies, namely when two clusters with comparable masses merge to form a
new more massive cluster. Aside from the obvious fact that these events 
are rather rare, one should also remember that the shock waves that  
develop during these events and that are responsible for both the heating
of the intracluster gas and the possible acceleration of cosmic rays, have 
relatively low Mach numbers \cite{gabici1} (e.g. $M\sim \sqrt{2}-2$). The 
spectrum of particles accelerated at these shocks is typically very steep 
\cite{gabici1}, and certainly quite steeper that $E^{-2.1}$, used in our 
calculations of the neutrino emissivity. Eq. \ref{eq:lumi} is therefore expected
to be an absolute upper limit to the cosmic ray luminosity (or equivalently
to the cosmic ray energy density averaged over the age of the cluster, 
$t_\tu{cl}\approx 10^{10} \rm yr$), in particular if used together with relatively 
flat spectra (e.g. $\propto E^{-2.1}$). These large energy densities would
also be in contradiction with the upper bounds to cosmic ray pressure found
in \cite{Reimer2004} for the Coma cluster, although these limits are somewhat 
model dependent. Lower cosmic ray luminosities and flatter injection spectra
are likely to be obtained if particle acceleration takes place at shock
waves that develop at the outskirts of clusters \cite{gabici2,gabici3,ryu},
due to their larger Mach numbers (namely lower temperatures of the background
unshocked gas).

\begin{figure}
  \centering
  \includegraphics[width=85truemm]{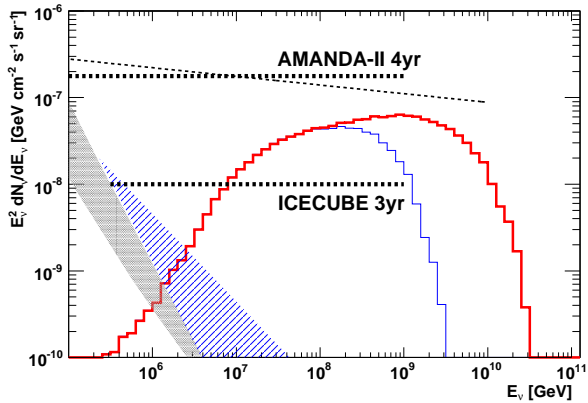}
  \caption{Upper limit to the neutrino flux from photopion production in
  clusters of galaxies for maximum energy of the protons
  $E_{max}=10^{20}$ eV (thick histogram) and $E_{max}=10^{19}$ eV (thin
  histogram). We also show the flux of atmospheric neutrinos (hatched
  region) including the contribution from prompt neutrinos (from
  \cite{atmo}). The sensitivities of AMANDA II and ICECUBE are also
  shown (from \cite{spiering}).}\label{fig:nudif} 
\end{figure}

The estimate of our upper limit on the diffuse neutrino flux from photopion 
production in clusters of galaxies is shown in Fig. \ref{fig:nudif}: the 
thick (thin) histogram represents our upper limit when the maximum energy 
of protons in clusters is $E_\tu{max}=10^{20}$ eV ($E_\tu{max}=10^{19}$ eV).
We also show there the upper bound on the neutrino flux from $pp$ interactions 
as found in \cite{bbp97} (dotted line). Note however that the upper limit 
in \cite{bbp97} was found by requiring that gamma rays generated in proton-proton 
inelastic scatterings in the ICM would saturate the EGRET diffuse gamma ray 
background. This would require $\xi\approx 0.3$ in Eq. \ref{eq:lumi} and 
injection spectrum $E^{-2.1}$.
As we already stressed, this situation is totally unphysical, as also 
discussed in \cite{bbp97}, but represents a solid upper limit to the flux
of neutrinos from clusters. 

In Fig. \ref{fig:nudif} the flux of atmospheric neutrinos is also plotted 
(from \cite{atmo}) in order to emphasize the flatness of the neutrino 
spectrum as predicted here and the energy region in which neutrinos from 
photopion production in clusters may overcome the atmospheric neutrino 
flux (the dark-hatched region is the standard atmospheric neutrino flux,
while the light-hatched region refers to prompt neutrino emission from charmed
mesons). This happens at energy $E\geq 5\times 10^6$ GeV. The upper limits 
that we show are still above the ICECUBE sensitivity (but below AMANDA II), 
as shown in the plot.

\begin{figure}
  \centering
  \includegraphics[width=85truemm]{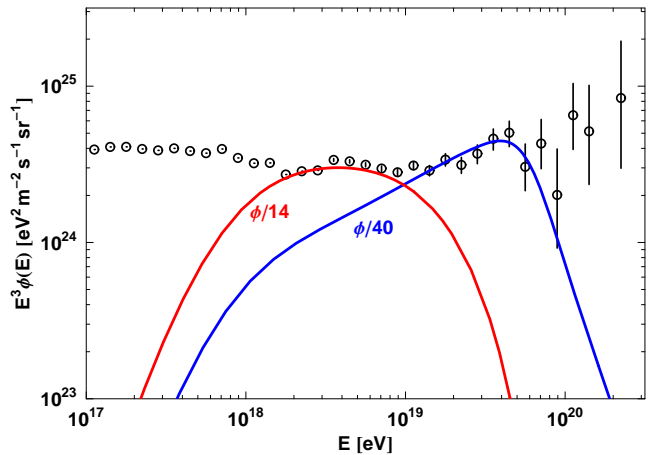}
  \caption{Flux of cosmic rays escaping clusters for $E_\tu{max}=10^{20}\eV$
  and $E_\tu{max}=10^{19}\eV$. The curves are normalized to the data of AGASA
  and Akeno. The factors 40 and 14 are the suppression factors needed to
  avoid overproduction of the extragalactic cosmic rays.}
\label{fig:escaped} 
\end{figure}

In obtaining the upper limit neutrino flux in Fig. \ref{fig:nudif}, we
did not consider however another important constraint: as we stressed in
Sec. \ref{sec:simple}, there are cosmic rays that are able to escape from 
the cluster volume in times shorter than the age of the cluster. Following
Fig. \ref{fig:alampg1} we adopt as a fiducial energy for the escape 
$10^{18}$ eV: particles with larger energies are not trapped inside the 
cluster and contribute to the flux of extragalactic cosmic rays at the 
Earth. It is important to realize that the spectrum of the escaping 
particles is very close to the injection spectrum at the source and 
does not resemble the spectrum of cosmic rays in the ICM (Eq. \ref{eq:equi}). 
The effect of energy losses on the spectrum of escaping cosmic rays is 
negligible. We calculated the flux of cosmic rays escaping clusters and
reaching the Earth following the well known approach to propagation 
illustrated in \cite{grigo}, which allows us to take into account 
photopion production, Bethe-Heitler pair production and adiabatic 
losses on cosmological scales. The results of our calculations are 
plotted in Fig. \ref{fig:escaped} for a maximum energy of the primary 
protons $10^{20}$ eV and $10^{19}$ eV. The low energy ($\sim 10^{18}$ eV)
cutoff in the spectra reflects the lack of confinement of higher energy 
cosmic rays. The curves are normalized to the data of AGASA and Akeno: 
in order to avoid exceeding the observed fluxes of cosmic rays at the 
Earth, the flux from clusters needs to be suppressed by 
a factor $\sim 40$ for $E_\tu{max}=10^{20}$ and by a factor $\sim 14$ for
$E_\tu{max}=10^{19}$. The neutrino fluxes in Fig. \ref{fig:nudif} need to be 
lowered by the same factors, so that the predicted neutrino fluxes from 
photopion production in clusters drop below the sensitivity limit for
ICECUBE. 

\section{Discussion and Conclusions}\label{sec:last}

We calculated the neutrino flux that could be expected from the photopion
interactions of very high energy protons with the infrared light and the
cosmic microwave background in clusters of galaxies. The main target for
these interactions is represented by the infrared photons generated by galaxies
in the cluster itself. This background can be much higher than the cosmic 
infrared background. This argument, added to the effective confinement of 
cosmic rays within the ICM for long times make clusters interesting targets 
for neutrino telescopes. 

The calculations have been carried out both analytically and with a
Monte Carlo code for the propagation and interaction of high energy protons.
There are many uncertainties involved in these calculations: the cosmic 
ray luminosity in the ICM averaged over cosmological times is very uncertain,
though somewhat limited by multifrequency observations of some clusters. 
These observations suggest that less than $1-10\%$ of the thermal energy 
content of cluster is in the form of cosmic rays. This limit is rather weakly
dependent on the diffusion coefficient, which instead strongly affects the 
escape times out of clusters.

The confinement time is important for our calculations because
neutrinos are mainly produced by cosmic rays with energy in excess 
of $10^{17}$ eV, where the confinement time is expected to be 
comparable with or shorter than the age of a cluster (Fig. \ref{fig:alampg1}).
The extent to which this is true depends on the
choice of the diffusion coefficient which is unknown and poorly
constrained due to our ignorance of the strength, structure and coherence
scale of the magnetic field in the ICM. As shown in Fig.~\ref{fig:alampg1}, 
for reasonable assumptions on this magnetic field, the diffusion coefficient 
in clusters is expected to have a dependence on momentum which is roughly 
linear in the energy region from $3\times 10^{18}$ to $3\times 10^{19}$ eV. 
We have checked the correctness of this finding in toy 
propagation models and concluded that without particle energy losses 
the containment time is exactly $E^{-1}$, despite the fact that the model 
of magnetic field turbulence adopted for the scattering of the particles is
of Kolmogorov type. While we believe that we describe well the proton 
propagation in the transition region, the diffusive part of
the propagation might be not fully realistic. If we overestimated the
 diffusion time of cosmic rays out of the clusters, then the neutrino flux is 
correspondingly reduced. This applies equally well to all neutrino flavors. 

The location where cosmic rays are injected also changes the
predicted neutrino fluxes but only in a rather marginal way. 
If the sources are spread over the cluster volume rather than 
being concentrated in the center as we assumed, then the neutrino
fluxes are estimated to be a factor $\sim 2$ lower than predicted here.
On the other hand it is reasonable to think that the IR background 
could have been higher in the past epochs due to luminosity evolution 
of the galaxies in clusters. This evolution might enhance the rate of 
neutrino production in the past, that would reflect into a higher 
diffuse neutrino flux for a given cosmic ray luminosity.

The neutrino fluxes from single galaxy clusters as calculated here 
were found to be undetectable with the currently planned and constructed 
detectors, such as IceCube~\cite{IceCube} and the European 
km$^3$ detector~\cite{km3}. Their detection would require 
the development of detection techniques able to provide much
higher detection volumes, such as the radio detection technique 
\cite{radio}.

The superposition of the tenuous neutrino fluxes from all clusters in
the universe generates a diffuse neutrino flux and turns out to be somewhat 
more interesting that the single sources. We estimated this diffuse neutrino
flux in Sec.~\ref{sec:diffuse} and our results are plotted in
Fig.~\ref{fig:nudif}. The upper limit to the neutrino flux appears to be
detectable by IceCube. On the other hand, the cosmic rays that are 
responsible for the production of these neutrinos also escape the
cluster and may reach us as cosmic rays. When the flux of these
cosmic rays is calculated, it exceeds the observed fluxes of cosmic
rays by a factor $\sim 40$ if the maximum energy of protons is $10^{20}$ eV
and by a factor $\sim 14$ if this maximum energy is $10^{19}$ eV. The
neutrino fluxes are expected to be suppressed by the same factors, unless
some other process inhibits the propagation of these cosmic rays on 
cosmological scales. One example of such processes, although not a viable
one, could be provided by the presence of a non negligible magnetic 
field in the intergalactic medium (spread over scales comparable with
the loss length of particles with energies $\sim 10^{19}$ eV). However, 
the value of the magnetic field needed to shrink the so-called magnetic 
horizon of particles of energy $\sim 10^{19}$ eV  to, say, 1/10 of the 
size of the universe is of the order of $\sim 10^{-8}\,\tu{G}$, which 
appears to be unreasonably large for the magnetic field averaged over 
the size of the universe. 

{\bf Acknowledgements} This work was supported in part by NASA grant 
 ATP03-0000-0080. The work of PB was supported through Cofin 2004.

\end{document}